\documentclass[aps,pre,twocolumn,showpacs,amsfonts,amssymb,floatfix,
superscriptaddress,nofootinbib]{revtex4-1}
\usepackage[utf8]{inputenc}
\usepackage{float}
\usepackage{graphicx}
\usepackage{amssymb}
\usepackage{amsmath}
\usepackage{bbm}
\usepackage{dcolumn}

\makeatletter

\newcount\hour \newcount\minute
\hour=\time \divide \hour by 60 \minute=\time \count99=\hour
\multiply \count99 by -60 \advance\minute by \count99 \hour=\time
\divide \hour by 60
\date{August 29, 2013} 

\makeatother

\usepackage{tabularx}
\begin{document}
%
\title[Self tolerance and idiotypic network]{Self tolerance in a minimal model
of the idiotypic network
}
\author{Robert Schulz}
\affiliation{Institute for Theoretical Physics, University Leipzig, POB
100 900, D-04009 Leipzig, Germany}

\author{Benjamin Werner}
\email{werner@evolbio.mpg.de}
\affiliation{Institute for Theoretical Physics, University Leipzig, POB
100 900, D-04009 Leipzig, Germany}
\affiliation{Present address: Max Planck Institute for Evolutionary Biology,
Pl\"on, Germany}

\author{Ulrich Behn}
\email{ulrich.behn@itp.uni-leipzig.de}
\affiliation{Institute for Theoretical Physics, University Leipzig, POB
100 900, D-04009 Leipzig, Germany}


\begin{abstract}
We consider the problem of self tolerance in the frame of a minimalistic model of the
idiotypic network. A node of this network represents a population of B lymphocytes of the
same idiotype which is encoded by a bit string. The links of the network connect nodes
with (nearly) complementary strings. The population of a node survives if the number of
occupied neighbours is not too small and not too large. There is an influx of lymphocytes
with random idiotype from the bone marrow. Previous investigations have shown that this
system evolves toward highly organized architectures, where the nodes can be classified
into groups according to their statistical properties. The building principles of these
architectures can be analytically described and the statistical results of simulations
agree very well with results of a modular mean field theory. In this paper we present
simulation results for the case that one or several nodes, playing the role of self, are
permanently occupied. We observe that the group structure of the architecture is very
similar to the case without self antigen, but organized such that the neighbours of the
self are only weakly occupied, thus providing self tolerance. We also treat this situation
in mean field theory which give results in good agreement with data from simulation.\\
\end{abstract}

\maketitle

\section{Introduction}

B lymphocytes express Y-shaped receptor molecules, antibodies, on their surface.
These antibodies have specific binding sites which determine their idiotype. All receptors of a
given B cell have the same idiotype. B cells with random idiotypes of remarkable
diversity are produced in the bone marrow. 

A B cell is stimulated to proliferate if its receptors are crosslinked by
complementary structures, unstimulated B cells die. Proliferation occurs if the
concentration of complementary structures is not too low or not too high, see
e.g. \cite{Coutinho89}. The latter condition refers to a steric hindrance for crosslinking if too many complementary molecules are around. Stimulating complementary
structures can be found on foreign antigens and on other, so-called
anti-idiotypic antibodies of complementary specifity. Thus B lymphocytes can stimulate each other and
form a functional network, the idiotypic network, as first proposed in
\cite{Jerne74}, see also \cite{Jerne84,Jerne85}.

The potential repertoire includes idiotypes that can recognize other complementary
structures, e.g. on the active sites of enzymes, hormones, and
neurotransmitters. Further there are idiotypic interactions of B
lymphocytes with T lymphocytes and between T cells.
Thus, the idiotypic network is not an autonomous entity of the adaptive immune
system, but is coupled to many other networks.

Even for an hypothetical autonomous B-lymphocyte system we have the requisites
of evolution, random innovation and selection.  So the architecture of the idiotypic network
can be conceived as the result of an evolution during the life time of an
individual.
In a revised version of the idiotypic network paradigm, the second generation idiotypic
network \cite{VC91,Coutinho03}, it was suggested that this architecture
comprises a strongly connected central part with autonomous dynamics and a hereto only sparsely connected periphery for
localized memory and adaptive immune response.

Already Jerne thought the idiotypic network to play an essential role in the control of
autoreactive idiotypes \cite{Jerne84}. Today the concept of idiotypic
networks is most popular in the research on autoimmune diseases, both in
theoretical studies and clinical context. Indeed, autoreactive antibodies are
regularly found in healthy individuals though in low concentrations. Antibodies
which escape other control mechanisms can be controlled by the idiotypic network
\cite{UWFC77}.
Anti-idiotypic antibodies specific to potentially autoreactive clones are found
in healthy individuals or in patients during remission, they are absent during periods of active autoimmune
disease \cite{Hampe12}. Autoimmune diseases can be related to perturbations of
the control of autoreactive clones
\cite{Avrameas91,SG97,Shoenfeld04,Pendergraftetal04,McGH05,TR10,RT10,Hampe12},
 as for example in Myasthenia gravis, a well known B-cell associated autoimmune
 disease \cite{DVK86}.

There are early attempts to model self tolerance and autoimmunity
mathematically within the network paradigm. In \cite{SV89} an idealized
architecture was proposed which comprises four groups of B-cell clones, a multi-affine group $A$, two mirror groups $B$ and $C$ with mutual coupling but no intra-group affinity, and a group $D$ which couples
with low affinity only to $A$. Based on this {\it ad hoc} architecture, 
computer simulations \cite{SVC89} and an analytical mean-field model
\cite{SvHB94} have shown the possible relevance for understanding
autoimmunity.

For more detailed accounts on the history of the paradigm, mathematical modeling
and new immunological and clinical developments the reader is referred to
\cite{Behn07,Behn11}.

In this paper we consider a minimal model of the idiotypic network proposed in
\cite{BB03} which describes the evolution towards complex, functional
architectures.
The most interesting architecture comprises densely linked core groups,  periphery groups,  groups
of suppressed clones, and groups of singletons which potentially interact only with the
suppressed clones. In the steady state, the size of these groups and their linking does
not change with time. The groups are build from clones of different idiotypes which share
certain statistical properties. The building principles of these architectures
can be described analytically \cite{SB06,STB12}, and the statistical
properties can be calculated in good agreement with simulations \cite{SB12}.
Here we investigate the evolution of the idiotypic network in the presence of self
toward an architecture where the expansion of autoreactive clones is controlled by idiotypic interactions.

The paper is organized as follows. In Section \ref{sec:model} we describe essential features of the model, its
update rules, the general building principles which allow to understand the structural properties of the expressed
networks architecture, and a tool which allows a real time identification of
patterns in simulations. In Section \ref{sec:MFT} we sketch the derivation of
the mean-field theory which allows to compute statistical properties if the structural properties of the pattern are known. In Section \ref{sec:IWS} we describe how the model should be modified in the presence of self. We report on simulations where for some protocols the impact of self is strong enough to
initiate a reorganization of the architecture such that the self is linked only to groups
with very low population. Results of a modified mean-field theory are in good agreement
with simulations. Finally, we give some conclusions and discuss problems for
further research.

\section{The model}
\label{sec:model}
In this paper we consider a minimal model of the idiotypic network
\cite{BB03} which is a coarse simplification of the real biological system
but retains most important features and reveals a surprising complexity. The model has only few parameters and allows an
analytical understanding of many of its properties.

A node $v$ of the network represents a clone of B lymphocytes of a given
idiotype together with its antibodies. The idiotype is encoded by a bitstring of
length $d$. The nodes are labeled by these bitstrings and can be conceived as
corners of a $d$-dimensional hypercube. Two nodes $v$ and $u$ are linked if
their bitstrings are complementary allowing for up to $m$ mismatches. The
corresponding undirected graph $G^{(m)}_{d}$ mimics the potential idiotype
repertoire of size $2^d$ with the possible interactions. Each node of the graph
is linked to $\kappa=\sum_{k=0}^{m} \binom{d}{k}$ neighbours. We only account
whether an idiotype is present or not, thus the corresponding node $v$ is either
occupied $n(v)=1$ or empty $n(v)=0$. The subgraph of occupied nodes with its links represents the expressed idiotypic network. The temporal evolution of the
network is induced by the following rules for parallel update:
\begin{enumerate}
\vspace{-1mm}
\item[(i)] Influx: Occupy empty nodes with probability $p$.\\[-5mm]
\item[(ii)] Window rule: Count the number of occupied neighbours $n(\partial v)$
of node $v$.
If $n(\partial v)$ is outside the window $[t_L, t_U]$, set the node $v$
empty.\\[-5mm]
\item[(iii)] Iterate.\\[-5.5mm]
\end{enumerate}
Extensive simulations have shown that the network evolves, depending
on the parameter choice, towards quasistationary states of possibly complex
architecture \cite{BB03}. This architecture is characterized by groups of
nodes that share statistical properties such as the mean occupation $\langle n(v) \rangle$ and
the mean occupation of neighbours $\langle n(\partial v) \rangle$. Here we
report on results for the following parameter setting. The length of the
bitstring is $d=12$, then the network has 4096 nodes. We allow $m=2$ mismatches
which makes the linking neither too sparse nor too dense, each node has $\kappa=
79$ neighbours.
The lower threshold $t_L$ of the window rule has its minimal nontrivial value $t_L=1$: for
survival of a clone the stimulation by at least one anti-idiotypic clone is
required. The upper threshold of the window rule is chosen as $t_U=10$, that excludes very
regular static patterns which are in our context not interesting, for more details, see
\cite{STB12}.
This parameter setting is best investigated. Simulations for longer bitstrings up to
$d=22$ have shown that many features are also found in larger networks and the major
concepts of structural analysis are still applicable \cite{HS12}.

There are general building principles of the network's architecture which have
been found by observing regularities in bitstrings that label nodes of the same
group \cite{SB06,STB12}. They make it possible to calculate the number of
groups, their size, and the linking between the groups. For a given architecture the nodes can be classified according to the entries in the
determinant positions of the bitstrings. If there are $d_M\leq d$ such
positions, the architecture has $d_M+1$ groups of size $\vert S_g \vert =
2^{d-d_M} \binom{d_M}{g-1}$, $g=1,\ldots,d_M+1$. Groups $S_g$ and $S_{d_M+2-g}$
have the same size. If the nodes of group $S_1$ have a string of determinant
bits with entries $\mathbf b_{d_M} \cdots \mathbf b_{d_1}$ in common, then the
nodes of group $S_2$ have one different entry in the determinant positions. In
general, nodes of group $S_g$ differ in $g-1$ positions from the determinant
bit entries of $S_1$.

The whole architecture can be build from smaller units, so called pattern
modules. These modules are the corners of a $d_M$-dimensional hypercube labeled
by the determinant bits, together with the allowed links. Since the number of
nondeterminant bits is $d-d_M$, the whole architecture is obtained by arranging
$2^{d-d_M}$ identical pattern modules and adding the allowed links between the
nodes of these modules.

We consider a pattern with $d_M$ determinant bits on a base graph $G^{(m)}_d$. The
elements $L_{ij}$ of the link matrix $\mathbbm L$ give the number of links of
any node in group $S_i$ to nodes in group $S_j$. Since the update rule counts
the number of occupied neighbours and all nodes of a group have the same
mean occupation these data are of obvious interest to formulate a mean-field
theory. A careful analysis of the bitstrings which encode the nodes of groups
$S_i$ and $S_j$ allows to give an explicit expression for $\mathbbm L$
\cite{SB06,STB12} which can be written as
\begin{align}
L_{ij} = & \sum \limits_{k=0}^{m} \sum \limits_{r=0}^{k} \binom{i-1}{r}
\binom{d_M-i+1}{j-1-r} \nonumber \\
 & \times \binom{d-d_M}{k+j-1-2r-(d_M-i+1)} .
\label{eq:linkmatrix}
\end{align}
Given a pattern with $d_M$ determinant bits there are $d_M+1$ groups,
therefore in Eq. \eqref{eq:linkmatrix} both $i$ and $j$ run from $1$ to $d_M+1$.
As every node has $\kappa$ neighbours, the row sum of $\mathbbm L$ yields
$\kappa$. Since $L_{ij}=L_{d_M+2-i,d_M+2-j}$ the link matrix is centrosymmetric, i.e. it
fulfills the identity $ \mathbbm L J = J \mathbbm L$ where the exchange matrix $J$ has
entries $1$ on the counterdiagonal and $0$ elsewhere. $\mathbbm L$
describes a directed graph.

\begin{figure}[h!!]
	\centering
	\includegraphics[width=85mm]{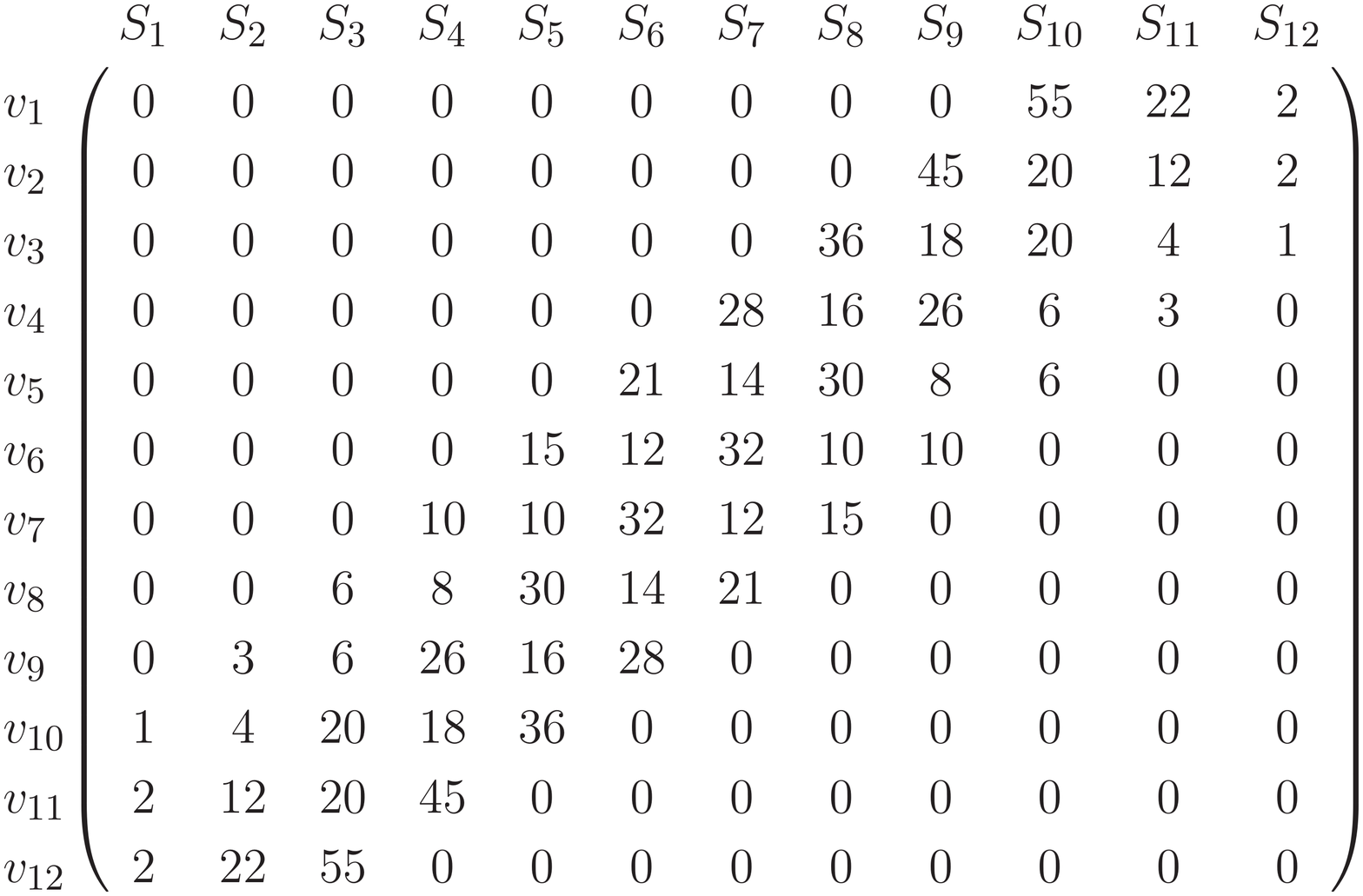}
	\caption{Link matrix for the 12-group architecture on
	$G^{(2)}_{12}$, described by Eq. \eqref{eq:linkmatrix}. The entries $L_{ij}$
	show the number of neighbours a node $v_i$ of group $S_i$ has in group $S_j$.
	Only the groups $S_6$ and $S_7$ have self coupling.}
\label{fig:SWBfig01}
\end{figure}

In Fig. \ref{fig:SWBfig01} we give the link matrix for the example of a 12-group
pattern which is studied in this paper.

In simulations, huge amounts of data are produced describing the occupation of
each of the $2^d$ nodes of the network in every single time step. An enormous reduction of
information can be reached by introducing a center of mass vector in dimension
$d$ which allows a real time identification of patterns and detection of pattern
changes \cite{STB12}. It is defined as
\begin{equation}
 \mathbf{R} = \frac{1}{n(G)} \sum_v n(v) \mathbf{r}(v) ,
 \label{eq:centerofmass}
\end{equation}
where the position vector $\mathbf{r}(v)$  of a node $v$  which is encoded by the
bitstring  $\mathbf{b}_d \mathbf{b}_{d-1} \cdots \mathbf{b}_1$ with
$\mathbf{b}_i \in \{0, 1\}$ has components $r_i(v) = 2 \mathbf{b}_i -1$. $n(G)$ is the
total occupation of the basegraph $G$. By definition, for a symmetrically
occupied base graph we have $\mathbf{R} = \mathbf{0}$, a symmetry breaking pattern is easy to identify.

\begin{figure}[h!!!]
	\centering
	\includegraphics[width=85mm]{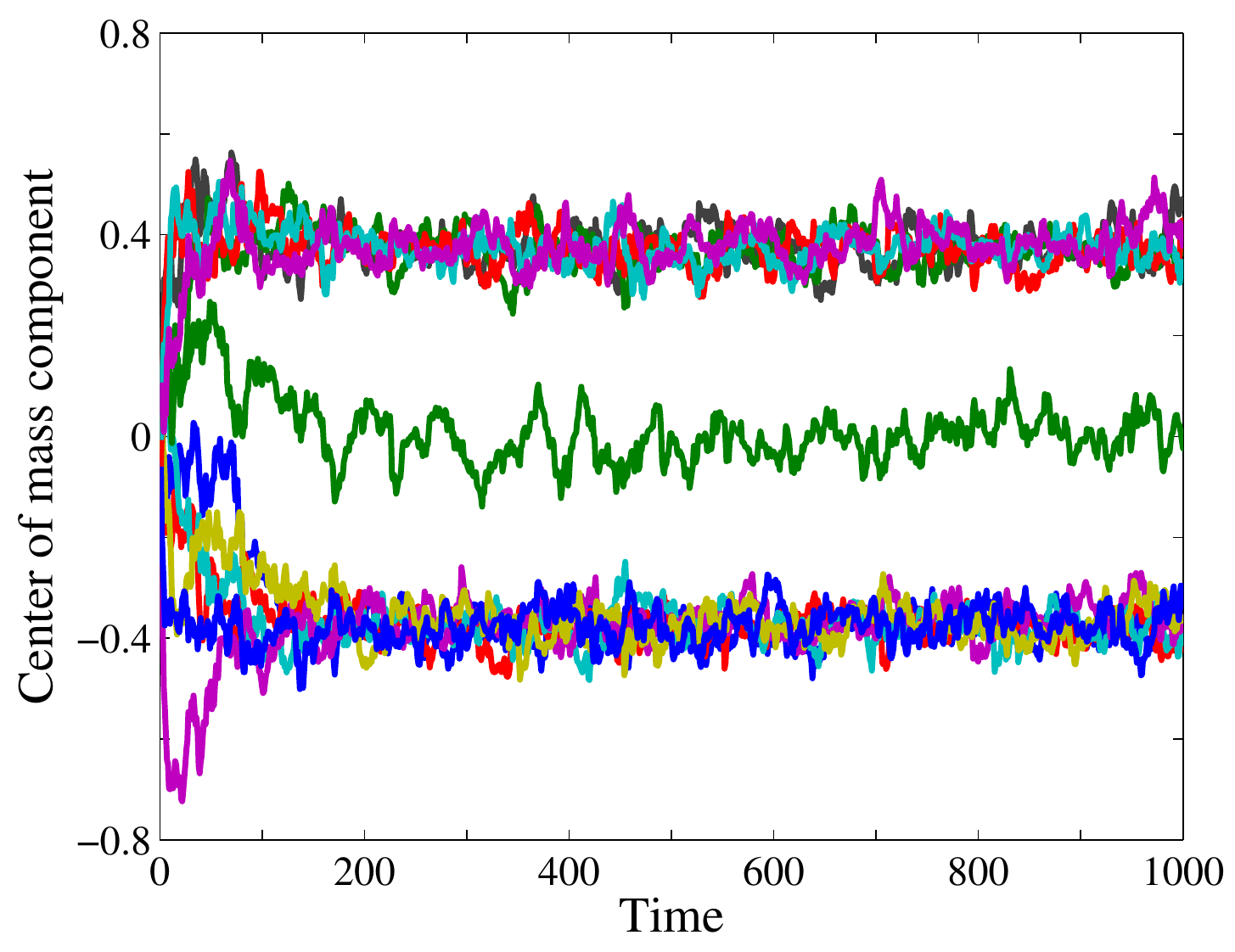}
\caption{Time series of
the center of mass vector components given by Eq. \eqref{eq:centerofmass}, here
on $G_{12}^{(2)}$ with $[t_L, t_U]=[1,10]$ for $p=0.074$. Every colour corresponds to one component of the
center of mass vector. We start from an empty base graph which is
gradually occupied, thus breaking the symmetry. A stationary state has evolved  after
about 200 time steps. Eleven components fluctuate around non-zero mean values, $R_1, R_3,
R_4, R_5, R_6, R_8 \approx -0.4$ and
$R_7, R_9, R_{10}, R_{11}, R_{12}\approx 0.4$, and $R_2$ fluctuates around zero.
Hence, as explained in the text, it is an 12-group architecture with
$d_M=11$ determinant
bits and all nodes in $S_1$ have a bitstring
$\mathbf{1111010000\cdot0}$, where the $\cdot$ represents the only
nondeterminant bit.}
\label{fig:SWBfig02}
\end{figure}

In Fig. \ref{fig:SWBfig02} we see the evolution towards a
$d_M=11$ pattern, where five determinant bits have the value one, another six
the value zero.
There is one nondeterminant bit which takes both
values zero and one. Supposed all nodes of a group are occupied with the same probability,
the contribution of each group to the nondeterminant component of $\mathbf R$ is
zero.

The procedure is fast, robust against defects of patterns, and allows to identify pattern
changes. Here we use this tool to characterize the behaviour of the network
if several nodes become permanently occupied to mimic the presence of self.
\section{Mean field theory}
\label{sec:MFT}
The existence of groups of nodes $S_g$ which share the mean occupation,
$\langle n(v_g) \rangle = n_g$ where $v_g \in S_g$, and the fact that the
window rule (ii) for update counts only the total of the occupied
neighbours suggest that a mean-field description of a given pattern, characterized by a pattern module of dimension
$d_M$ and the corresponding link matrix $\mathbbm L$, is possible. Indeed, the
modular mean-field theory developed in \cite{SB12} delivers results in good
agreement with simulations. We shortly sketch the derivation here to make the
modifications understandable which are necessary when modeling the presence of
self.

Application of the update rules to a state characterized by $\mathbf
n~=~(n_1,\cdots,n_{d_M+1})^{\text{T}}$ leads to a new state $\mathbf n'$ given
by
\begin{align}
\mathbf n' = \mathbf f(\mathbf n) ,
\label{eq:fixed-point-eq}
\end{align}
where the nonlinear function $\mathbf f$ depends on the update rules and on the
pattern we want to describe. We know that a node $v_g$ of group $S_g$ has
$L_{gl}$ neighbours in $S_l$. If the mean occupation in $S_l$ is $n_l$, the new
mean occupation after the influx with probability $p$ is $\tilde n_l = n_l +
p(1-n_l)$. The probability that $k_l$ nodes of the neighbourhood in $S_l$ are
occupied after the influx is
\begin{align}
\binom{L_{gl}}{k_l} \;
\tilde n_{l}^{\;k_l} (1- \tilde n_{l} )^{L_{gl}-k_l} .
\label{eq:binomial}
\end{align}
Supposing that the groups are independent, the probability that for a
microconfiguration with fixed $k_l$, $l=1,\ldots,d_{M}+1$, a total of
$\sum_{l=1}^{d_M+1} k_l$ neighbours is occupied is simply the product of factors
\eqref{eq:binomial} for each group. Summing over all microconfigurations and taking into account the window rule leads to
\begin{align}
& \left[ \sum \limits_{k_l=0}^{L_{gl}} \right]_{l=1}^{d_M+1}
\mathbbm{1}(t_{L} \leq \sum \limits_{l=1}^{d_M+1} k_{l} \leq t_{U}) \nonumber \\
 & \times \prod \limits_{l=1}^{d_M+1} \binom{L_{gl}}{k_l}
\tilde n_{l}^{k_l} (1- \tilde n_{l} )^{L_{gl}-k_l} ,
\end{align}
where the indicator function $\mathbbm 1(\cdot)$ gives one, when the window rule
in the parentheses is fulfilled, otherwise zero. The last result should be
multiplied with the mean occupation of a node of the considered group after the
influx $\tilde n_g = n_g + p(1-n_g)$ which gives
\begin{align}
n'_{g} = & \tilde n_{g} \left[ \sum \limits_{k_l=0}^{L_{gl}}
\right]_{l=1}^{d_M+1}
\mathbbm{1}(t_{L} \leq \sum \limits_{l=1}^{d_M+1} k_{l} \leq t_{U}) \nonumber \\
 & \times \prod \limits_{l=1}^{d_M+1} \binom{L_{gl}}{k_l}
\tilde n_{l}^{k_l} (1- \tilde n_{l} )^{L_{gl}-k_l} .
\label{eq:MFT}
\end{align}
Iterating Eq. \eqref{eq:MFT}, for $g=1,\ldots,d_M+1$, the $\mathbf n'$ converge
to a fixed point $\mathbf n^\star$.
Since $\mathbf f(\mathbf n)$ is a nonlinear function, several fixed points may
exist. As a thumb rule it is sufficient to start close to the values seen in
simulations, to reach the 'right' fixed point.
\section{Idiotypic network and self}
\label{sec:IWS}
\begin{figure}[h!!]
	\centering
	\includegraphics[width=85mm]{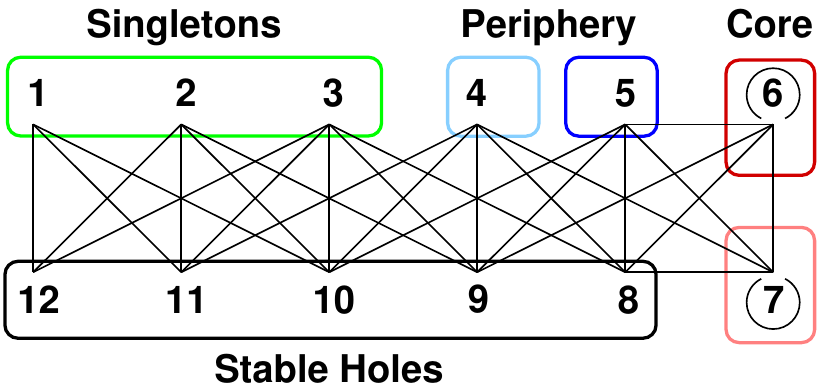}\\[5mm]
\caption{The network on $G^{(2)}_{12}$ develops for a range of $p$
from $0.026$ to $0.078$ a steady state architecture of 12 groups which build a
weakly occupied densely connected core (red), a periphery (blue) which is highly occupied and couples to the core and to the
group of stable holes (black), and a group of singletons (green) which only couple to the
stable holes.}
\label{fig:SWBfig03}
\end{figure}

The most interesting architecture which has been found in extensive simulations
on the base graph $G^{(2)}_{12}$ for $[t_L,t_U]=[1,10]$ for a range of $p$ from
$0.026$ to $0.078$ is the 12-group architecture. The groups comprise two self
coupled core groups, two peripheral groups which couple only to the core and
five groups of stable holes. Stable holes are typically unoccupied since their
occupied neighbours exceed $t_U$. Finally, there are three groups of singletons
which are neighboured only by stable holes. A scheme of these architecture is
given in Fig. \ref{fig:SWBfig03}.
Singletons have an occupation of
$0.2$ to $0.8$, depending on $p$, the population of the densely linked core
groups is kept below $0.07$, and the holes are almost empty, for details see
Fig. 8 in \cite{SB12}.
This architecture strongly resembles the central and peripheral parts of the
second generation idiotypic network \cite{VC91}.

The simplest possible way we can imagine to mimic the presence of self is to
permanently occupy one or several nodes of the base graph and to see what architecture
evolves. Naturally the impact of the permanently occupied nodes, the
self, increases with their number. The impact also depends on the influx rate
$p$ since the 12-group architecture becomes unstable for $p \gtrapprox 0.08$.
Close to this threshold the strongest impact is to be expected. 

\subsection{Simulations}
\begin{figure}[h!]
	\centering
	\includegraphics[width=85mm]{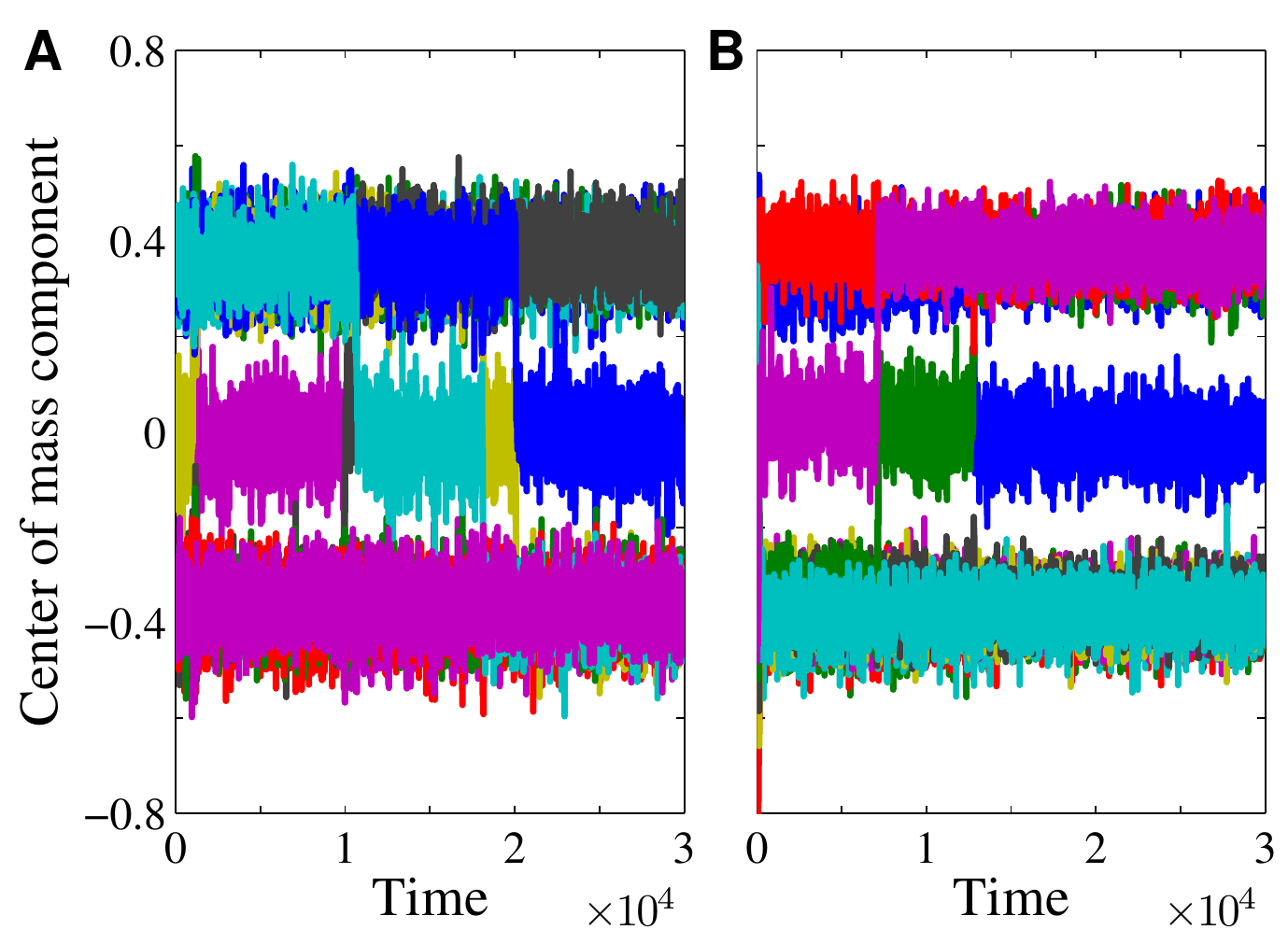}
\caption{Evolution of the idiotypic network with self. The figure
	displays time series of the center of mass vector components for an established
	12-group architecture when at $t=0$ nodes of the hole group $S_{10}$ are
	permanently occupied. {\bf(A)} For $p=0.076$ one node of $S_{10}$ is
	permanently occupied. After five changes of the determinant bits, which can be
	easily identified by the colour changes of the trajectory fluctuating around
	zero, indicating severe reorganization of the pattern, the node belongs to
	the periphery group $S_5$ and the architecture is stable from thereon. {\bf(B)} For $p=0.074$ ten nodes of $S_{10}$ are
	permanently occupied. After few changes of the determinant bits a stationary
	state has evolved where four self nodes belong to the singleton group $S_3$ and
	six self nodes to the periphery $S_5$.}
	\label{fig:SWBfig04}
\end{figure}

We performed extensive simulations for different protocols. Here we
describe only few most instructive cases.
We permanently occupy one node of the hole group $S_{10}$ of an established 
12-group pattern for $p=0.076$. The hole groups have many occupied neighbours and
a self node staying there would be subject of a heavy autoimmune response.
After few iterations the former stable pattern destabilizes under the
presence of the self and collapses.
Thereafter a new 12-group architecture evolves where the self node is now located in a group with
only weakly occupied neighbours, which could be one of the singleton or periphery
groups. Fig. \ref{fig:SWBfig04}A shows an example where the self is
finally in the peripheral group $S_5$.

If we permanently occupy more than one node the
scenario is similar. Fig. \ref{fig:SWBfig04}B shows an example where we
have permanently occupied 10 nodes of the hole group $S_{10}$ of an established
12-group pattern for $p=0.074$. The reorganization of the architecture is faster
and in the new steady state the self nodes are found in singleton and periphery
groups.
Starting from an empty base graph with several permanently occupied nodes, one
observes that the architecture evolves from the very beginning such that the
self nodes have only weakly occupied neighbours and thus are tolerated.

We also performed simulations where for a stationary 12-group pattern all
members of the hole group $S_{10}$ are permanently occupied. After
reorganization of the architecture, in the steady state all self nodes belong to
singletons and periphery groups and are never seen in a core or a hole group. If
we start from an established 12-group pattern and permanently occupy one of the
singletons or periphery groups this state will be stable for very long periods
of time.
\subsection{Mean field theory with self}
It is possible to modify the mean-field theory to describe a stationary
architecture in the presence of self. 
We thus can describe situations where in an established pattern nodes are
permanently occupied and the impact is so small that no reorganization sets in.
If the impact is strong enough that a reorganization occurs and a new steady
state emerges, we also can describe the statistical properties of this steady
state provided that we know its architecture. 

We first consider one permanently occupied node of group $S_s$.
It can be directly seen by nodes of group $S_g$ if $L_{sg}>0$. The group $S_g$
contains $L_{sg}$ nodes that see the self. For these nodes we should modify the
mean-field mapping Eq. \eqref{eq:MFT}.
The node of $S_s$ which is permanently occupied should be exempted from the
combinatorics of possible and allowed micro-configurations.
Thus we need to replace $L_{gs}$ by $L_{gs}-1$. Observe that
$\binom{L_{gs}-1}{k_s}$ in the modified Eq. \eqref{eq:MFT} is zero if $L_{gs}-1$
is smaller than zero or $k_s$. To account for the permanently occupied self node we should decrease both
thresholds of the window condition by 1. For the $\vert S_g \vert - L_{sg}$
nodes of $S_g$ which do not see the self node, the mapping is not modified. For example,
for an influx with $p=0.07$ and one permanently occupied node in a hole group
or in a core group, $\langle n(\partial v) \rangle$ increases by about 1 and
$\langle n(v) \rangle$ decreases by about $20\%$ if $v$ sees the self directly.
The mean-field theory agrees with the simulation within $3\%$ to $5\%$.

The case that all nodes of a group $S_s$ are permanently occupied is even
simpler because all nodes in group $S_g$ see the same number $L_{sg}$ of self nodes. We
only have to modify the window condition decreasing both thresholds by $L_{sg}$.
Note that if $t_U-L_{sg}<0$ the modified window condition cannot be fulfilled
and the indicator function $\mathbbm 1(\cdot)$ in the modified Eq.
\eqref{eq:MFT} returns 0.
Tab. \ref{Tab:01} gives a detailed comparison of simulation and mean-field
theory for the case that all 110 nodes of the singleton group $S_{10}$ are occupied for
$p=0.074$.

\newcolumntype{.}{D{.}{.}{-1}}
\begin{table*}[t!!!]
	\centering
	\begin{ruledtabular}
	\begin{tabular}{c|..|..|..|..}
& 	\multicolumn{4}{c|}{$\langle n(v) \rangle$}& \multicolumn{4}{c}{$\langle
n(\partial v) \rangle$}\\[2mm]

Group & \multicolumn{2}{c|}{Simulation} & \multicolumn{2}{c|}{MFT}
&\multicolumn{2}{c|}{Simulation} & \multicolumn{2}{c}{MFT} \\[2mm] \hline
            &        &        &       &        &       &      &     & \\[-2mm]
	$S_{1}$ & 0.0 &  & 0.0   & & 71.75 &(54.01) & 71.45 & (53.99)\\ 
	$S_{2}$ & 0.0 &  & 0.0   & & 60.34 &(53.86) & 60.29 &(53.96)\\
	$S_{3}$ & 0.0 &   & 0.0   & & 59.62 &(53.50) & 59.85 &(53.52)\\
	$S_{4}$ & 0.0 &   & 0.0   & & 36.70 &(34.86) & 36.62 &(34.72)\\
	$S_{5}$ & 0.0 &   & 0.0   & & 31.5 &(29.62)  & 31.6 &(29.7)\\
	$S_{6}$ & 0.002 &(0.001) & 0.0   & & 13.52 &(13.53) & 13.61 &(13.63) \\
	$S_{7}$ & 0.01 & & 0.003 & & 10.12 &(10.09) & 10.10 &\\
	$S_{8}$ & 0.677 &(0.675) & 0.671 & & 0.15 &(0.14)  & 0.07 & \\
	$S_{9}$ & 0.706 &(0.695) & 0.682 &(0.683) & 0.025 &(0.018) & 0.01 & \\
	$S_{10}$ & 1.0 &(0.685) & 1.0 &(0.692)  & 0.02 &(0.0) & 0.01 &\\
	$S_{11}$ & 0.685 &(0.684) & 0.691 &(0.685) & 0.001 &(0.0) & 0.0 &  \\
	$S_{12}$ & 0.685  &(0.682) & 0.692 &(0.685) & 0.001 &(0.0) & 0.0 &  \\
	\end{tabular}\\[5mm]
	\end{ruledtabular}	
 \caption{12-group pattern with self after rearrangement, see Fig.
 \ref{fig:SWBfig05}.
 The 110 nodes of the singleton group $S_{10}$ are permanently occupied to mimic the presence of self antigen. The
table shows the mean occupation $\langle n(v) \rangle$ and the mean occupation of neighbours $\langle
n(\partial v) \rangle$ for all groups as obtained for $p=0.074$ from simulations and from
mean-field theory (MFT) with a $d_M=11$ module. When deviating, the data for the
case without self are given in parentheses. The groups $S_1, \dots, S_{5}$ have direct
neighbours in $S_{10}$, where $S_{1}$ has the most ones. Therefore, the change
in $\langle n(\partial v) \rangle$ due to self is largest for $S_{1}$. Results
from simulation and mean-field theory are in good agreement.}
\label{Tab:01}
\end{table*}

For $N_s$ self nodes with $1<N_s<\vert S_s \vert$ the modification is also
possible but more intricate and will not be reported here.

\begin{figure}[h!!!!]
	\centering
	\includegraphics[width=85mm,height=82mm]{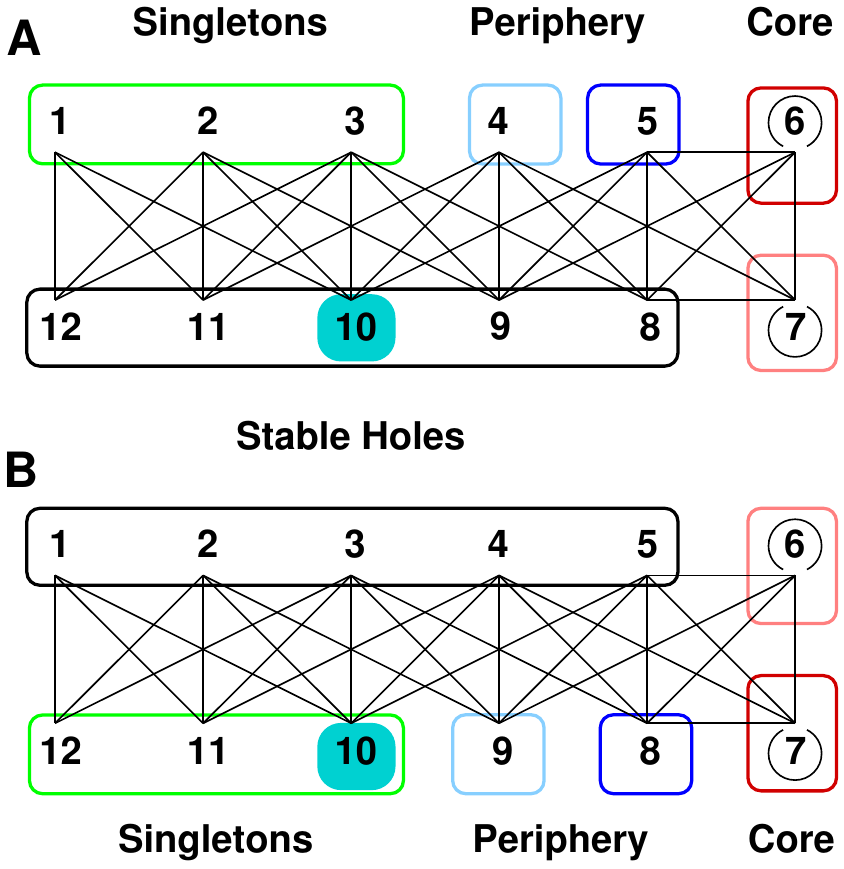}\\[5mm]
\caption{12-group architecture with self. {\bf (A)} We permanently
occupy one of the hole groups, group 10 (cyan), thus mimicking the permanent presence of self. This state is not favourable since
the self couples to singletons and periphery which have a high occupation. {\bf
(B)} Letting the thus prepared system evolve, it soon reaches a new steady
state, still a 12-group architecture, but organized such that the self now belongs to the singletons and
thus couples only to the almost empty stable holes. The self-recognizing
idiotypes are controlled by the network, thus providing self tolerance.}
\label{fig:SWBfig05}
\end{figure}

\begin{figure}[h!!!!]
	\centering
	\includegraphics[width=85mm,height=60mm]{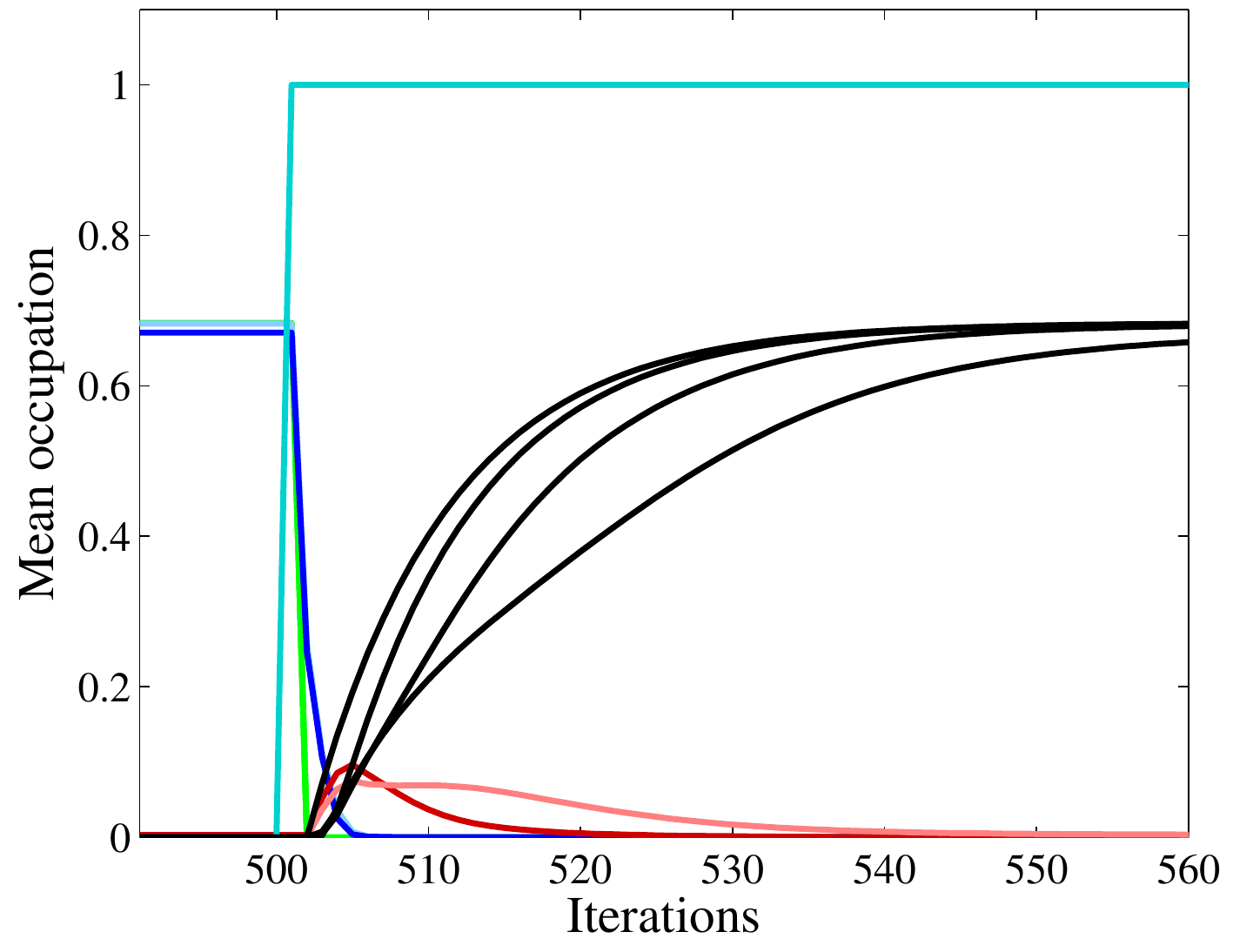}\\[0mm]
\caption{Mean field theory of a 12-group architecture with self. The
figure shows the mean occupation of the 12 groups as obtained from iterating Eq. \eqref{eq:MFT} for
$p=0.074$. We start with the autonomous system where in the steady state 
singletons (green) and periphery (blue) have  a mean occupation per node
$\langle n \rangle \approx 0.68$, whereas core (red) and stable nodes have
$\langle n \rangle \approx 0$. At iteration step 500 we permanently occupy the
110 nodes of the hole group $S_{10}$ (cyan). The fixed point of Eq.
\eqref{eq:MFT} looses its stability and a new mirrored architecture emerges where the
permanently occupied nodes, the self, now belong to the singletons which have
neighbours only in the empty hole groups, cf. Fig. \ref{fig:SWBfig05}.
The occupation of the previous singleton and periphery groups drops down to almost zero, whereas
the previous hole groups become occupied as typical for singletons and
periphery. After an temporary increase the core groups return to its previous
occupation. For a detailed comparison of mean-field theory and simulation see
Tab. \ref{Tab:01}.}
\label{fig:SWBfig06}
\end{figure}

Encouraged by the good quantitative agreement between the steady states obtained
in simulations and mean-field theory we also looked at the time series of
$\mathbf n$ generated by the mean-field mapping for a $d_M=11$ pattern at $p=0.074$. We
start with the fixed point $\mathbf n^{\star}$ which describes a 12-group
pattern where the groups are ordered as in Fig. \ref{fig:SWBfig05}A.
At iteration step 500 we permanently occupy the hole group $S_{10}$. The time
series, cf. Fig. \ref{fig:SWBfig06}, shows that this state immediately destabilizes and that a
reorganization sets in.
The pattern converges to a new state where the self
belongs to the new singleton group $S_{10}$. These singletons have only
neighbours in the new unoccupied hole groups, see Fig. \ref{fig:SWBfig05}B. The
network controls the expansion of the autoreactive idiotypes in the hole groups -- thus providing
self tolerance.

We note in this context that due to the centrosymmetry of
the link matrix of the autonomous network without self, given a
fixed point
$\mathbf n^{\star} =
(n^{\star}_1,n^{\star}_2,\ldots,n^{\star}_{d_M+1})^{\text{T}}$,
there exists always a mirrored fixed point $\mathbf
n^{\star}_{\text{mirror}} = (n^{\star}_{d_M+1},\ldots,n^{\star}_2,n^{\star}_{1})^{\text{T}}$.
Obviously this symmetry is broken if self is present. 
\section{Conclusion}
\label{sec:Conclusion}
For a minimal model of the idiotypic network in the presence of self, we have
reported on simulation results for several protocols where, if the impact of
self is sufficiently strong, the network evolves toward an architecture that
controls the expansion of self-reactive clones thus providing self tolerance.
For the simplest cases that only one node or all nodes of a group are
permanently occupied we have modified the mean-field theory and find good
agreement of analytical and simulation results.

The network in the presence of self has been previously studied in
simulations for one self node on the base graph $G^{(3)}_{12}$ with
weighted links. The weights were given according to the number of
mismatches of the linked nodes and the window condition was modified
accordingly. The patterns are slightly easier to destabilize which explains why
the phenomenon of self tolerance was first observed in that version of the
model \cite{Wer10}.

Further studies should systematically explore the system's behaviour for other
protocols, e.g. for arbitrary numbers of self nodes possibly distributed over
the whole base graph, desirably in both simulations and an accordingly
extended mean-field approach.

It is of obvious interest to investigate in the frame of the model possible
reasons for failure of self tolerance. Transitions from a healthy self-tolerant
state to an autoimmune state by a perturbation, possibly an
ordinary infection, of the clones that control the autoreactive idiotypes should
be considered, together with the reverse phenomenon of 'spontaneous' remission
from an autoimmune to a healthy state. Therapeutic strategies adopting the network
paradigm \cite{FS05} which consist in stimulating the
protective clones that control the autoreactive clones, instead of applying
immunosuppressive drugs, could be modeled.

From the viewpoint of statistical physics or systems biology the question
appears natural and most interesting whether there is a general principle which
guides the evolution of the idiotypic network.
\section*{Acknowledgement}
The authors thank Holger Schmidtchen and R\"udiger K\"ursten for valuable
discussions.
%

%
%
%
\end{document}